\documentclass[11pt]{article}
\usepackage{graphicx}
\usepackage{amssymb}

\usepackage{amssymb}
\usepackage{amsmath}

\renewcommand{\H}[1]{\ensuremath{\mathbf{H}(#1)}}

\def\BEn{\begin{enumerate}}
\def\EEn{\end{enumerate}}

\def\from{\kern-2pt\leftarrow\kern-2pt}
\def\breve{\mathaccent"7014 }

\def\U{\breve}

\def\0#1{{\mathrm{#1}}}
\def\1#1{{\mathbb{#1}}}
\def\2#1{{\mathbf{#1}}}
\def\3#1{{\mathcal {#1}}}
\def\4#1{{\mathsf{#1}}}
\def\5#1{{\mathfrak{#1}}}
\def\6#1{\overline{#1}}
\def\7#1{{\check{#1}}}
\def\8#1{{\widehat{#1}}}
\def\9#1{{\breve{#1}}}
\def\Ui{\9{\imath}}

\def\wh{\widehat}

\def\dag{\dagger}

\def\dag{\dagger}

\def\wh{\widehat}
\def\U{\breve}
\def\Ui{\U{\imath}}

\def\bra{\langle}

\def\ket{\rangle}

\def\ox{\otimes}

\def\x{\times}

\def\BE{\begin{equation}}
\def\EE{\end{equation}}
\def\BEA{\begin{eqnarray}}
\def\EEA{\end{eqnarray}}

\def\rar{\rightarrow}

\begin{document}

\title{\huge Finite Quantum Kinematics of the Harmonic Oscillator\footnote
{Based on the Ph. D. Thesis of M. Shiri-Garakani\cite{SHIRI}.} }
\author {{\bf \large
Mohsen Shiri-Garakani\footnote{To whom correspondence should be
addressed at Department of Chemistry and Physical Sciences, Pace
University, Pleasantville, NY. e-mail:
\texttt{mshirigarakani@pace.edu}}
 \quad   David Ritz Finkelstein\footnote{School of Physics,
Georgia Institute of Technology, Atlanta, GA. e-mail:
 \texttt{df4@mail.gatech.edu}}
}\\
}

\date{July 2004}

\maketitle
\abstract{
Arbitrarily small changes in the commutation relations suffice
to transform
 the usual singular quantum theories into regular quantum theories.
 This process is an extension of canonical quantization that we call general quantization.
 Here we apply general quantization to the time-independent linear harmonic oscillator.
The unstable Heisenberg group
becomes the stable
group SO(3).
This freezes out the zero-point energy of
very soft or very hard oscillators, like those responsible for the
infrared or ultraviolet divergencies of usual field theories,
without much changing the medium oscillators.
It produces pronounced violations of equipartition and
of the usual uncertainty relations for soft or hard oscillators,
and  interactions between the previously uncoupled
excitation quanta of the oscillator,
weakly attractive for medium quanta, strongly repulsive
for soft or hard quanta.
}

\section{Make it simple}

The three main evolutions of physics in the twentieth century have a
suggestive family resemblance.
Each introduced a new kind of
non-commutativity.
The new non-commutativity in special
relativity was that of boosts,
in
general
relativity and the Standard Model gauge theories that of
infinitesimal translations, and in
quantum theory that of
filter operations.
The seminal work of Segal \cite{SEGAL},
which stimulated the present work,
pointed out
that further changes of this kind are necessary for stability
and suggested one.
Our main goal is finiteness, not stability,
but
the stabilizing changes Segal suggested
lead ultimately to
a finite quantum theory, including  one of space-time.
Such a
theory has been sought by physicists
since the formulation of quantum theory.

By gently modifying the commutation relations of
an existing quantum theory one
produces a simpler theory
with
the existing quantum theory as a suitable limiting case,
and with nearly  the
same continuous symmetries.
A special form of
Segal's general concept  was  applied
 retroactively to the relation between special relativity and Galileo relativity
\cite{Inonu1952, Inonu1964}.
More proactively,  Snyder's space-time quantization  \cite{SNYDER} was an attempted regularization and moved unwittingly toward simplicity but did not simplify the momentum algebra.
Vilela made the first efforts to
found a new particle theory on a simple algebra \cite{VILELA}.
Something like
the  regularization of the harmonic oscillator proposed by Segal
is now under study by several groups from several  points of view
 \cite{Carlen2001, Atakishiyev2003,  KUZMICH, MANFREDI,THOOFT}.

For example, 'tHooft \cite{THOOFT} studies  the classical particles
on a circle and shows that under certain conditions, this system is
equivalent to a quantum harmonic oscillator. The  work of  Vilela Mendes
differs from others in presenting
the quantum harmonic oscillator as a limit case of a ``more
quantum" oscillator that has a more stable algebra in the sense of
Segal.
We follow that line here.

 Naturally one discards the unregularized theory in favor of the
regularized one. This last step is  overlooked in some older
studies.

General quantization changes the quantization rules.
It replaces the usual quantization prescription by
the following one:

{\em Make the commutator algebra of the generators a simple Lie algebra.}

Briefly put:  If the algebra is simple, keep it simple. If it is not simple, make it simple
by the least change possible.

We apply this strategy to all the algebras postulated in a physical theory,
on the grounds that they must all depend continuously on experimental data
that are subject to error.
An algebra that is not stable is not empirical but is at least partly based on ideology.
It is not possible to eliminate all such ideology-based hypotheses from physics.
But it is possible to reduce their number systematically.

This implies that canonical quantization and special and general relativization
are relatively small parts of a vast unifying drift toward simplicity and unity of the groups
beneath our physical theories.
General quantization is an attempt to take part in that drift.

In the case of the algebra of variables, semisimplicity is as good as simplicity,
since it implies a direct sum that we can reduce to a simple
term forever by a single measurement of the superselection variable
that distinguishes these terms.

Canonical quantization introduced one quantum constant and stabilized
the atom.
General quantization introduces as many quantum constants as necessary
to stabilize the group of the theory.

General quantization has no effect on (say) the rotator quantum algebra
 with given angular momentum $l$
 ($\2L \x \2L=\2L$,  $\2L\cdot \2L=l(l+1)$),
whose group is already isimple.
But it changes the quantum dynamics of a free particle deeply,
revising  the theories of both space and time.

We test general quantization
here on the kinematics of the linear harmonic oscillator,
a ubiquitous constituent of all present field theories,
and compare
the finite quantum oscillators with the usual quantum oscillators,
which are singular in several senses.
The differences have profound consequences for extreme energy physics:
the physics of both very high and very low energies.

Planck's quantum constant $h$  froze out the very stiff oscillators
responsible for the infinite heat capacity of cavity radiation
in Maxwell's theory, but
the zero-point energy of the resulting quantum theory
of electromagnetism still diverged, however,
unless one arbitrarily
replaced the local Lagrangian and Hamiltonian of Maxwell
by non-local ones tailored to have some finite  zero-point energy,
usually set to  0 on grounds of Lorentz invariance.
Indeed, the quantum theory of the harmonic oscillator carries the germs of all
the main divergences of quantum field theory.
Its basic operators of position $q$, momentum $p$
and  Hamiltonian $H\sim \frac 12 (p^2+ q^2)$
are undefined on almost every vector $\psi$ in its Hilbert space:
$q\psi=p\psi=H\psi=\infty$.
Such divergences occur in a quantum theory if and only if
its Hilbert space is infinite-dimensional.

The usual quantum oscillator theory is also unstable in the Segal sense detailed below.
It
is not as unstable as the classical theory, which
has the operators $q$ and $p$ commute,
and we have become accustomed to its foibles,
but it is still not operational, in that its
basic
operations usually cannot be carried out.
General quantization
makes its Hilbert space finite-dimensional.
The result is a finite quantum  theory whose operations
can in principle be carried out,
with
two  Segal quantum constants $h', h''$ besides
the usual Planck constant.

We find that introducing these constants
freezes out even the offending zero-point oscillations
of extremely hard or soft oscillators
without
greatly changing the zero-point energies of medium ones.
The frozen oscillators have infinitesimal zero-point energies
compared to the usual quantum values.
They also grossly violate
the usual equipartition and uncertainty relations.

This toy model illustrates how a finite quantum theory of the cavity
might  produce a finite zero-point energy without conflicting with
the many finite predictions and symmetries of the usual quantum theory.
We propose that  the  linear harmonic
field oscillators considered fundamental in present quantum physics
--  those of supposedly fundamental fields, not those of elastic solids, say ---
are actually finite quantum oscillators
near the bottoms of their energy spectra.
The unobserved oscillators responsible for
the infrared and ultraviolet divergencies
of present quantum theories
are frozen by finite quantum effects described here
and contribute negligibly to the zero-point energy.

The change we carry out here is not enough to make quantum field theory finite.
For that we must also simplify the Heisenberg algebra of
the space-time operators $x^{\mu}$ and $\partial_{\mu}$.
This replaces the manifold theory of space-time, which assumes an infinity of events,
by a simple quantum theory,
with only a finite number of disjoint events,
though the number may be arbitrarily large.
Field theory has compound algebra on two levels,
that of the underlying space-time and that of the overlying
canonical commutation relations.
In this paper we change only one level.

\section{Algebra flexing and flattening.}
A {\it
semi-simple} group is  a Lie group whose Lie algebra
has no invariant solvable subalgebras;
a Lie algebra $A$ being solvable if for some integer $n\ge 0$\/,
$ A^n=\{0\}$.
Then its Lie algebra has no radical.
A group that is not semisimple we call {\it compound}\/.
General quantization reduces the radical
and ultimately eliminates it.

Lie products $\times$ on a given vector space $A$,
also called structure tensors,
 form
a submanifold $\{\times\}$ of the tensor space $A\ox A\dag\ox A \dag$.
A  {\it regular} (stable, robust) algebra is one
that is unchanged  up to
isomorphism by all sufficiently small changes
 in its structure tensor (Lie product) within the manifold $\{\times\}$.
For example,
the Lorentz algebra is stable against corrections to the speed of
light.
By  algebra {\it flexing} we mean a homotopy of the structure tensor
of a compound algebra that makes it semisimple.
Algebra {\it flattening} is  the inverse process.
The well-known contraction
process of In\"on\"u and Wigner  \cite{Inonu1952} is a
special case of
 flattening
accomplished by a one-parameter group of dilations of
 a coordinate system of the Lie algebra in a fixed direction.
 The
inverse to group contraction is {\it group expansion}
\cite{Gilmore1994} and is a special case of flexing.

Several regularization
processes have been used to remove unwanted infinities from
physics.
Unphysical regularizations
cope with the divergencies of a  theory
without changing the finite results.
They are regarded as giving the theory meaning rather
than changing the theory.
These include
 Pauli-Villars
regularization \cite{Pauli1949}, lattice regularization (lattice
gauge theory) \cite{Wilson1974} and dimensional regularization
\cite{Bollini1964, Bollini1972, THOOFT}.
They contain regularization parameters
that go to  singular limits,
like the lattice spacing going to 0.
Physical regularizations, on the contrary,
 change the finite predictions as well as making the infinite ones finite,
 and are intended as distinct physical theories in their own right.
Their regularization parameters  do not go to a singular limit
but must be determined by experiment.
 The most famous example  is Planck's, which ultimately led to quantum theory.
This was a simplification in that the associative algebra generated
by the position and momentum variables
has a Lie algebra with an infinite-dimensional radical  for $h=0$
but only a
one-dimensional radical for $h>0$.

Physical regularizations
are subtler  than  unphysical ones
but their consequences for human thought
have been more
dramatic.
General quantization is proposed as a physical regularization.

Compound algebras are {\it unstable} with respect to a small change in
their structures \cite{SEGAL}.
Flexing stabilizes them.

Conversely,
flattening destabilizes.
Approximating  a circle  by a  tangent line or  a sphere by a
tangent plane are well-known flattenings.
The
circle and sphere are  finite and their flattened form  is infinite.
Finite dimensional representations  of the group of the sphere
---
such as spherical harmonic polynomials
---
form a complete set on the sphere,
and all the operators of an irreducible representation have finite
bounded spectra.
On the other hand the tangent plane is not compact
and requires infinite-dimensional representations of the translation
group for a complete set,
and its algebra generators have unbounded
spectra.

\section{Simplifying the Heisenberg Algebra}

The Heisenberg Lie algebra \H{1} is
defined by the  commutation relations:
\BEA
[p,x]&=&-i\hbar \cr
[i,x]&=&0\cr
[x,i]&=&0 \label{eq:Heisenberg2}
\EEA
It is compound and
the imaginary unit $i$
generates its radical.
Segal proposed to simplify $\0H{1}$ by introducing two more quantum
constants, which we designate here by  $\hbar'$ and $\hbar''$.
His expanded commutation
relations are, except for notation,
\BEA
[p,x]&=&-\hbar i \cr
[i,p]&=&-\hbar' x\cr
[x,i]&=&\hbar''p\label{eq:IHeisenberg}
 \EEA
  \cite{Carlen2001,
VILELA,SEGAL}.
The irreducible unitary representations of this group are
infinite-dimensional.
To avoid possible
divergences and other problems, we instead use the
$\0S\0O(3)$
regularization \cite{Baugh2001,Baugh2003,Finkelstein2003,Galiautdinov2002}\BEA
[p,x]&=&-i\hbar \cr
[i,p]&=&-x\hbar'\cr
[x,i]&=&-p\hbar''\label{eq:SHeisenberg}
 \EEA
Ultimately we will need an indefinite metric for physical reasons,
but not for the time-independent harmonic oscillator.

Regularizing the Heisenberg algebra means changing the
role of $i$ in the theory from constant central element to
quantum variable operator on the same footing as $p$ and
$q$. We call this $i$-activation.
The new variable that it introduces is called a regulator.
A previous exploration in quaternion quantum theory
activated an $i$ that served as the electromagnetic axis $\eta(x)$
that resolves the
electroweak gauge boson into electromagnetic and weak
bosons \cite{Finkelstein1963},
and
gives mass to the charged
partner of the photon through
the St\"{u}ckelberg-Higgs effect.
This led to a natural $\0S\0U(2)$ that was interpreted
as isospin.
That theory was dropped because it did not
leave room for
color $\0S\0U(3)$.
Here we activate $i$ on more
principled grounds, namely the principle of simplicity.
There is now
plenty of room for internal groups like color $\0S\0U(3)$,
though they do not arise for the harmonic oscillator.

General quantization leads to the same kind of factor-ordering problems
as the special case of canonical quantization.
To reduce  these we regularize not Hermitian observables directly but
 skew-Hermition
generators
\BE \wh{q}=iq,\qquad \wh{p}=-ip.   \EE
The usual quantum commutation relations are then
\BEA [ \wh{q}, \wh{p} ] & = &\hbar {i}\cr
[ {i} , \wh{q}]     & = & 0,    \cr
[ \wh{p},{i} ]       & = & 0,   \cr
i^2 & = & -1.
\EEA
The regularized
generators $\U{q},\U{p},\Ui$ obey
\BEA
[\U{q}, \U{p}]&=&\hbar \Ui,\cr
[\Ui,\U{q}]&=&\hbar' \U{p}, \cr
[\U{p},\Ui]&=&\hbar''\U{q},
\label{eq:XQCR}
 \EEA
We suppose  $\hbar, \hbar', \hbar''
>0$ so the orthogonal group is $\0S\0O(3)$.
The quantities with a breve ``\ $\breve{}$\ '' are the new expanded
quantum operators.
In this way the simplification process introduces a new dynamically
variable generator $\Ui$, somewhat as general-relativization
introduced the new dynamical variable $g_{\mu\nu}$\, the
gravitational metric tensor field.
The most primitive theory with a
dynamical variable like $\Ui$ is quaternion quantum field theory
\cite{Finkelstein1963}. There $\Ui$ generates  rotations about the
electric (or electromagnetic) axis in isospin space, defining a
natural Higgs field. We suppose that the present generator $\Ui$ is
also a Higgs field.

When general quantization introduces new group generators in this way
for simplification, we call these regularization operators
or ``regulators.''
The physical constants to which regulators reduce
in the singular theory we call regularization constants  or ``regulants.''
Examples of a regulator in present physics are
 the Riemann curvature
of space-time (as the commutator of covariant transports)
and  the gravitational field itself
(as the anticommutator of unit Clifford vectors).
Examples of regulants are $\hbar$ and $c$.

Except for scale factors the simplified commutation relations are
those  of an $\0S\0O(3)$ quantum
skew-angular-momentum operator-valued vector
$\U{\2L}=\U{\2L}\times
\U{\2L}$ for a dipole rotator in three dimensions.
We assume an
irreducible representation with
\BE \U{\2L}^2=-l(l+1)
\EE
where $l$ can have any non-negative half-integer eigenvalue.
In the present work it suffices to consider only integer values of $l$.
Then the
$\U
L_x, \U L_y, \U L_z$  are
represented by $(2l+1)\times (2l+1)$ matrices obeying
\BEA
[\U{L}_1,\U{L}_2]&=&\U{L}_3, \cr [\U{L}_2,
\U{L}_3]&=&\U{L}_1,\cr [\U{L}_3,\U{L}_1]&=&\U{L}_2,
\cr (\U{L}_1)^2+(\U{L}_2)^2+(\U{L}_3)^2&=&-l(l+1)\/.
\label{eq:LCR}
\EEA
We fix the scale factors with
  \BEA
\U{q}&=&Q \U{L}_1, \cr \U{p}&=&P\U{L}_2,\cr \Ui&=&
J \U{L}_3,
 \label{eq:scales} \EEA
By  (\ref{eq:XQCR})
\BEA J&=&\sqrt{\hbar'\hbar''} \;=\; 1/l, \cr
Q&=&\sqrt{\hbar\hbar'},\cr
P&=&\sqrt{\hbar \hbar ''}.
\label{eq:JQP}
\EEA
 The commutation relations (\ref{eq:LCR}) and the angular momentum quantum number $l$
 determine a simple (associative) enveloping
algebra Alg$(\2L,l)$
The
spectral spacing of the $\U{L}_3$ is 1, so the finite quantum
constants $Q,P, J$ serve as quanta
   of position, momentum and $\Ui $.
Since $q, p$ are supposed to have continuous spectra in quantum
theory, the constants $Q, P$ must be very small on the ordinary quantum
scale.
It follows that $J=QP/\hbar$ is also very small
on that scale and $l\gg 1$.

For $l\gg \sqrt{l}\gg 1$,
variations $\delta (\Ui^2) \le O(l^{-1/2}) \ll
1$ about $(\Ui)^2=-1$ can be negligible at the same time as the
spectral intervals $\delta p\le  P\sqrt{l}$ and $\delta q\le
Q\sqrt{l}$ for  quasicontinuous $p, q\approx  0$.
This simulates the usual oscillator.

\section{Finite Linear Harmonic Oscillator}\label{sec:XQHO}

Now we specialize to the oscillator by fixing a Hamiltonian.
For given finite-quantum constants $P, Q$ the finite harmonic
oscillator has a Hamiltonian of the form
 \BE \label{eq:FQHO}
H=\frac{P^2L^2_x}{2m}+\frac{kQ^2L^2_y}{2}:= \frac K2\left ( L_x\/^2
+ \kappa^2 L_y\/^2\right)
 \EE
 where
  \BE
K=\frac{P^2}{m},\quad \kappa^2=\frac{\hbar'mk}{\hbar''}.
\EE
For fixed $\hbar, \hbar', \hbar''$, all  finite oscillators  are divided
into three kinds with ill defined boundaries:
{\em medium}\/,  where kinetic and potential terms in $H$ are of
comparable size ($\kappa \sim 1$);  {\em soft}\/, when the potential energy
term is dominant ($\kappa\to 0 $); and hard, when the kinetic energy
term is dominant ($\kappa \to \infty$).
Examination of the Hamiltonian of a spin-zero scalar field
(Klein-Gordon field) in quantum field theory shows that the
possibilities $\kappa\ll 1$ and $\kappa\gg 1$ are also important.
The  oscillators that
give rise to infrared divergencies of the quantum field theory
correspond to soft oscillators of the finite quantum theory.
Those that
feed ultraviolet divergencies
correspond to hard oscillators.

\section{Medium oscillators}
The  case $\kappa=1$
is  symmetric under rotations about the $z$ axis, and so is especially simple
\cite{THOOFT}.
Since
\BE
 (\U{L}_1)^2+(\U{L}_2)^2+(\U{L}_3)^2=
(\U{L}^2),
\EE
\BE
\U{H} = \frac{K}{2}\left(l(l+1)+(\U{L}_3)
^2\right)\label{eq:Hamiltonian}
\EE
The oscillator quantum
number $n$ that labels the energy level is now
\BE n=l+m.
\EE
The expanded energy spectrum
is
\BE E_n= \frac{K}{2}\left(l(l+1)-(n-l) ^2\right) = {lK}\left(n +
\frac{1}{2} -\frac{n^2}{2l}\right) \EE
For $n\ll\sqrt{l}\ll l$ this
reproduces the usual uniformly-spaced oscillator energy spectrum as closely as
desired, but with multiplicity  2 for each level instead of 1.

The ground-state energy for this oscillator is
\BE E_0 =
\frac{1}{2}Kl=\frac 12 1/2 \hbar\omega,
\label{eq:groundenergy}\EE
exactly the usual
oscillator ground energy,
since $Kl=\hbar\omega$.

The main new feature is that this finite oscillator has an upper energy limit
\BE E_{max}= \frac{1}{2}Kl(l+1)\EE
 as required by  a finite quantum theory.

In the general case of $\kappa \sim 1$ we obtain
an upper bound for the ground energy by a variational
approximation with the  trial function $|L_z=\pm l\ket$.
This
reproduces our previous result (\ref{eq:groundenergy}),  now
as an upper bound for the ground energy of a
medium FLHO:
\BE E_0\leq \frac{1}{2}Kl.\EE
Medium oscillators
have many states with $m$-value close to its extremum value $m=\pm l$.
The usual
Heisenberg uncertainty principle
\BE(\Delta p)^2(\Delta
q)^2\geqslant\frac{1}{4}\bra i [p,q]\ket^2=\frac{\hbar^2}{4}.
\EE
becomes
\BE
 (\Delta L_x)^2(\Delta L_y)^2\geqslant
\frac{\hbar^2}{4}\bra L_z\ket^2_{|L_z\approx\pm l\ket} \EE
for a low-lying energy level of a medium
oscillator.
By (\ref{eq:scales}) and (\ref{eq:JQP}),
 \BE
  (\Delta p)^2(\Delta y)^2\geqslant  \frac{\hbar^2}{4}\EE
  for large $l$.
  So
medium oscillator states in low-lying energy levels have uncertainties at or
above the lower limit set by the Heisenberg uncertainty principle.

\section{Soft  oscillators}

Recall our finite quantum oscillator Hamiltonian
\BE \U{H} =
\frac{K}{2}\left( \U{L}^2_x + \kappa^2 \U{L}^2_y\right) \EE
When $\kappa \ll 1$
we  can estimate the spectrum  of $\U H$ using
perturbation theory.
The unperturbed Hamiltonian for our problem is
the kinetic energy
\BE
H_0=\frac{K}{2}L^2_x
 \EE
 and the unperturbed eigenvectors are $|L_x=m\ket
$
 so the unperturbed energy levels are
\BE
E_m(0)=\frac K2 m^2.
\EE
The first-order shifts
are
\BE \delta E_m=\frac K2 \bra L_x=m| L^2_y|L_x=m\ket. \EE
Due
to the axial symmetry of $|L_x=m\ket
$,
\BE \bra L_x=m| L^2_y|L_x=m\ket = \bra L_x= m|
L^2_z|L_x=m\ket . \EE
Therefore the energy shift is,  to lowest order in $\kappa^2$,
 \BEA
 \frac K2  \bra L_x=m| \kappa^2 L^2_y|L_x=m\ket&=&
 \frac K 4  \kappa^2 \bra  m| L^2_x+L^2_y|m\ket \cr
&=&\frac K 4 \kappa^2 \bra m| L^2-L^2_z|m\ket\cr
&=&\frac K 4 \kappa^2  l(l+1)-m^2 \EEA
The new energy spectrum is then
\BEA
 E_m&\approx &\frac{K}{2}m^2+\Delta E_m\nonumber
\\&=&\frac{K}{2}m^2+ \frac{1}{4}K\
\kappa^2\left[l(l+1)-m^2\right]\label{eq:SoftFLHO}\EEA
The estimated upper
bound for the energy is
\BE
E_{max}\approx \frac{1}{2}Kl^2(1+\frac{\kappa^2}{2l})\EE

For $\kappa \rar 0$ this reproduces the upper bound for the
unperturbed hamiltonian $L^2_z$, as it should.
The zero-point energy $E_0$ of first-order perturbation theory is
\BE E_0\approx \frac 14 \kappa^2 K l(l+1) \EE
For $\kappa \to 0$ this is infinitesimal compared to the usual QLHO.

 A soft oscillator  shows no resemblance to the usual
quantum oscillator.
Its  energy levels  do not have  uniform
spacing.
Its  kinetic energy
dwarfs its potential energy,
so equipartition is grossly violated.
 The low energy states
are near $|L_x=0\ket$ instead of $|L_z=\pm l\ket$.
Its  $p$
degree of freedom is frozen out. It is ``too
soft to oscillate:"
There is not enough energy in the $q$ degree of
freedom, even at its maximum excitation, to produce one quantum of $p$.
The uncertainty relation reads
 \BE
 (\Delta L_x)^2(\Delta L_y)^2\geqslant
\frac{\hbar^2}{4}\bra L_z\ket^2_{|L_x\approx 0\ket}\approx 0 \EE
Therefore \BE \Delta p \Delta q \ll \frac{\hbar}2, \EE
which violates the Heisenberg uncertainty principle grossly.

\section{Hard  oscillators}

The story is just reversed for hard  oscillators but the gross violations of usual quantum principles remain the same.
A hard oscillator has much greater potential than kinetic energy.
Its low energy states are now near
$|L_y=0\ket$ instead of $|L_z=\pm l\ket$ (the medium case) or $|L_x=0\ket$
(the soft case).
Its  $q$ degree of
freedom is frozen out. It is ``too hard to
oscillate.'' There is not enough energy in the $p$ degree of freedom,
even at maximum excitation, to arouse one quantum of $q$.

A hard oscillator can also be treated by perturbation theory.
The
kinetic energy is the perturbation.
We may carry all the of the main results in the previous section for
soft FLHO oscillators to the hard ones simply by replacing $\kappa$
with $1/\kappa$ and $K$ with $K\kappa^2$.
A hard FLHO shows no resemblance to the
usual QLHO.
Its zero-point energy $E_0$  is now
\BE E_0\approx \frac K4   l(l+1) \EE
For $\kappa \to \infty$ this is infinitesimal compared to the usual
quantum oscillator zero-point energy.
Its energy levels of a hard oscillator are not
uniformly spaced.
Its  uncertainty relation reads
 \BE
 (\Delta L_x)^2(\Delta L_y)^2\geqslant
\frac{\hbar^2}{4}\bra L_z\ket^2_{|L_y\approx 0\ket}\approx 0 \EE
Therefore
\BE \Delta p \Delta q \ll \frac{\hbar}2, \EE
which seriously violates the Heisenberg uncertainty principle again.

\section{Unitary Representations}

Variables $p$ and $q$ do not have finite-dimensional unitary
representations in classical and quantum physics. They are
continuous variables and generate unbounded translations of each other.
But since in the
finite quantum theory, all operators become finite and quantized, we
expect all translations to become rotations with simple
finite-dimensional unitary representations.

The canonical group of a classical oscillator becomes
the unitary group of an infinite-dimensional
 Hilbert space for a quantum oscillator,
and the unitary group of a $2l+1$ dimensional Hilbert space
for the finite oscillator.

The Lie algebra generated by momentum and position
as infinitesimal symmetry generators is
$\2H(1)$ for the classical and quantum oscillator
and  the $\2S\2O(3)$ angular momentum algebra
for the finite oscillator.
The corresponding Lie groups are the Heisenberg group $\2H(1)$
and $\2S\2O(3)$.

Unitarily inequivalent unitary representations of the canonical
 commutation relations
are forbidden in quantum mechanics but present and important in  quantum field theory,
but general quantization eliminates them.
After general quantization the  commutation relations  become  those of a large simple group,
and we presently explore the orthogonal groups.
Once its invariants are fixed, as by measurement,
the finite-dimensional unitary representations of this group are uniquely determined up to unitary equivalence.
Yet the general quantized theory approaches the usual singular theory
in an appropriate limit,
where the dimension of the representations grow without bound and the group becomes compound.
The inequivalent representations of quantum field theory must return
in that singular limit.
More than that we cannot say at this stage in the development.

\section{Conclusion}\label{sec:discussion}

We suggest that algebra flattening causes the
infinities of present physics.
Since quantum theory began as a regularization procedure of Planck,
it is rather widely accepted that further regularization of present
quantum physics calls for further quantization,
but what to quantize
and how to quantize it remains at least a bit unclear.
If we regard quantization as another step in group regularization,
the rest of the path becomes clear.
It is
blazed with radicals  ripe for
relativization.
General quantization  of the
linear harmonic oscillator  results in a finite quantum theory
with three quantum constants $\hbar, \hbar', \hbar''$ instead of the
usual one.
The finite quantum oscillator
 is isomorphic to a dipole rotator with
$N=l(l+1)\sim 1/(\hbar'\hbar'')\gg 1$ states and bounded
Hamiltonian $H= A
(L_x)^2 + B(L_y)^2$.
Its position and momentum variables are
quantized with uniformly spaced bounded
finite spectra
and supposedly universal quanta of position and momentum.
For fixed quantum constants and large $N\gg 1$ there
are three broad classes of finite oscillator,
 soft, medium, and hard.
The field
oscillators responsible for infra-red and ultraviolet divergences
are soft and hard respectively.
Medium oscillators have  $\sim \sqrt N$ low-lying states
having nearly the same zero-point energy and level spacing as the
quantum oscillator and nearly obeying the Heisenberg uncertainty principle and
the equipartition principle.
The corresponding rotators are nearly
polarized along the $z$ axis with $L_z\sim \pm l$.

The soft and hard
oscillators  have infinitesimal 0-point energy,
and grossly violate
both equipartition and  the Heisenberg uncertainty relation.
They do not resemble the quantum oscillator at all.
Their low-lying
energy states correspond to rotators with  $L_x\sim 0 $ or $L_y\sim
0$ instead of  $L_z\sim \pm l$.
Soft oscillators have frozen
momentum $p\approx 0$ because their maximum potential energy is too small to
produce one quantum of momentum.
Hard oscillators have frozen
position $q\approx 0$  because their maximum kinetic energy is too small to
produce one quantum of position.

The zero-point energy of a physical oscillator likely contributes to its
gravitational field.
 It will be interesting to estimate its
contribution to astronomical gravitational fields.
 For a consistent
estimate  we should regularize the quantum field theory,
not just one of its oscillators.
 This changes not only the structure of the
 individual oscillators, as considered here,
 but also the number and distribution of the
 oscillators.  We leave this study for later,
 but it is already easy to say how it will proceed,
 and what form it  will take.

Field theory has compound algebras on two levels, that of the
underlying space-time and that of the overlying canonical
commutation relations.
regularization. In this paper we change only the top  level,
but
to simply field theory we must also simplify
the Heisenberg algebra of the lower-level space-time operators $x^{\mu}$ and
$\partial_{\mu}$.
This replaces the manifold theory of space-time,
which assumes an infinity of events, by a simple quantum space-time
theory, with only a finite number of disjoint events, though that
number may be arbitrarily large.

We must then combine two finite-dimensional algebras,
that of the local field variables
nd that of the space-time-energy-momentum variables,
to make the finite-dimensional algebra of the field theory.
In c discrete theories, the combination process
is exponentiation $S^T$ where $S$ is the local field-variable state-set
and $T$ is the space-time set.
In the q/c theories that work best today,
where the numerator q indicates that $S$ is quantum
and the denominator $c$ indicates that $T$ is still classical,
an exponential still exists and is used.
General quantization leads to q/q theories.
In that case the usual exponential $S^T$ becomes basis-dependent,
and the most economical invariant construct that includes
all the special cases is the exterior algebra over $S\ox T$,
but this is still finite-dimensional.
Since general quantization gives time too a beginning and an end, the
time-development is certainly not unitary.
As a result the problem of reconciling  unitarity, causality, and
Lorentz invariance \cite{Finkelstein1969} is eliminated.
On the other hand,  since the Lorentz
group is already simple, Lorentz invariance is unaffected by general quantization.

General quantization
 modifies low- and high-energy physics.
 Because the low-lying energy levels
 of medium oscillators have nearly uniform spacing,
 the energy of two excitations is but slightly less
 than the sum of their separate energies.
 The corresponding quanta nearly do not interact,
and the small interaction that they have is attractive.
 For soft or hard oscillators,
 the energy level varies quadratically
 with the energy quantum number.
 The energy of two quanta of oscillation is twice the sum of their separate energies, for example.
 The corresponding quanta have a repulsive interaction of great strength;
 the interaction energy is equal to the total energy of the separate quanta.
 Thus the simplest regularization leads to
 interactions between the previously uncoupled
excitation quanta of the oscillator,
weakly attractive for medium quanta, strongly repulsive
for soft or hard quanta.

Like Dirac's theory of the ``anomalous''
magnetic moment of the
relativistic electron,
 these extreme-energy effects depend on factor ordering.
They can be adjusted to fit the data by re-ordering factors and so
are not crucial tests of the theory.
A group regularization of a time-dependent free Dirac equation
 has been carried out \cite{Galiautdinov2002} and the
extension to interactions is under study.

\section{Acknowledgments}
We thank James Baugh, Eric Carlen, Heinrich Saller, John Wood, and an anonymous referee for helpful discussions.

\tableofcontents
\end{document}